\documentclass[aps, prd, twocolumn]{revtex4-2}

\usepackage{graphicx}
\usepackage{dcolumn}
\usepackage{subcaption}
\usepackage{bm}
\usepackage{braket}
\usepackage{xcolor}

\begin{document}
\preprint{APS/123-QED}

\title{Revisiting the evidence for precession in GW200129 with machine learning noise mitigation}
\author{Ronaldas Macas}
\affiliation{University of Portsmouth, \relax Portsmouth, PO1 3FX, UK}
\author{Andrew Lundgren}
\email{Andrew.Lundgren@port.ac.uk}
\affiliation{University of Portsmouth, Portsmouth, PO1 3FX, UK}
\author{Gregory Ashton}
\affiliation{Department of Physics, Royal Holloway, University of London, TW20 0EX, UK}

\begin{abstract}
GW200129 is claimed to be the first-ever observation of the spin-disk orbital precession detected with gravitational waves (GWs) from an individual binary system.
However, this claim warrants a cautious evaluation because the GW event coincided with a broadband noise disturbance in LIGO Livingston caused by the 45\,MHz electro-optic modulator system.
In this paper, we present a state-of-the-art neural network that is able to model and mitigate the broadband noise from the LIGO Livingston interferometer.
We also demonstrate that our neural network mitigates the noise better than the algorithm used by the LIGO-Virgo-KAGRA collaboration.
Finally, we re-analyse GW200129 with the improved data quality compared to the data used by the LIGO-Virgo-KAGRA collaboration and show that the evidence for precession is still observed.
\end{abstract}

\maketitle

\section{Introduction}
\label{sec:intro}
The gravitational-wave (GW) era started in 2015 when a binary black hole (BBH) merger was detected~\cite{gw150914}.
Since then, the LIGO-Virgo-KAGRA (LVK) collaboration observed binary neutron star and neutron star-black hole systems, as well as many other BBH mergers~\citep{o1_o2_cat, o3a_cat, o3b_cat}.

GW analyses often assume that the noise is Gaussian.
While this is true in many cases, about 24\% of the GW candidates from the third LVK observing run (O3) were flagged as having a non-Gaussian noise nearby~\citep{o3a_cat, o3b_cat}.
If the non-Gaussian noise is not accounted for, GW searches and source parameter estimation can be affected, including the sky localisation which is used for the electromagnetic follow-up efforts~\citep{glitch_pe_1, glitch_pe_2, glitch_pe_3, gw170817_bw, skyloc}.

One of the GW candidate events from O3, GW200129, attracted interest because it is claimed to be the first-ever individual BBH that has strong evidence for spin-induced orbital precession~\cite{Hannam_2022}.
Initial analysis performed by the LVK collaboration with two waveform models found evidence for precession when using IMRPhenomXPHM but not with SEOBNRv4~\cite{o3b_cat}. Ref.\,\cite{o3b_cat} gave equal weight to both analyses, thus leaving the properties of GW200129 unclear.
In addition, \citet{Nitz_2023} found GW200129 to be precessing when analysed with the IMRPhenomX waveform model~\cite{Pratten_2020}.
\citet{Hannam_2022} put forward an analysis based on the NRSurrogate waveform model that is more faithful to numerical simulations~\cite{Varma:2019csw} and found the orbital precession of GW200129 to be 10 orders of magnitude higher than any previous weak-field measurement.

The precession claim warrants a cautious evaluation because GW200129 coincided with excess noise at the Livingston interferometer caused by the $45$\,MHz electro-optic modulator system~\citep{Davis_2022, rf_noise}.
Luckily, this noise was also recorded by multiple witness channels, namely \texttt{LSC-POP\_A\_RF45\_I\_ERR\_DQ}, \texttt{LSC-POP\_A\_RF45\_Q\_ERR\_DQ} and \texttt{LSC-POP\_A\_RF9\_I\_ERR\_DQ} which we refer to as \texttt{RF45-I}, \texttt{RF45-Q} and \texttt{RF9-I}, respectively.

The LIGO-Virgo-KAGRA collaboration used \texttt{gwsubtract} to model and mitigate the excess noise in the GW strain channel \texttt{h(t)}~\cite{o3b_cat}.
The \texttt{gwsubtract} algorithm was first developed to remove the `jitter' noise from LIGO Hanford interferometer during the second LIGO-Virgo observing run, which resulted in 30\% improvement in LIGO Hanford sensitivity~\cite{davis_linsub}. 
The algorithm was similarly applied for noise mitigation around GW200129 for the Livingston data: it estimated the coupling between the \texttt{RF9-I} and \texttt{h(t)} channels, and removed the predicted excess noise from \texttt{h(t)}.
This cleaned version of the Livingston data was then used in the LVK and Ref.\,\cite{Hannam_2022} analyses.

The implementation of \texttt{gwsubtract} for GW200129 and the algorithm itself has multiple shortcomings.
Firstly, the derivation of \texttt{gwsubtract} relies on the assumptions that the data is Gaussian and stationary~\cite{davis_linsub}, while the excess power in the Livingston data is non-Gaussian by definition. 
Furthermore, the algorithm assumes that the noise coupling must be linear which is not necessarily the case for GW200129.
Finally, the LVK collaboration used only a single channel, \texttt{RF9-I}, to estimate the noise coupling, even though \texttt{RF45-I} and \texttt{RF45-Q} also recorded the noise.

Additional studies suggest that the evidence for precession using the \texttt{gwsubtract} data is robust when using a NRSurogate waveform model~\citep{islam2023, Payne_2022}.
However, it is unclear if the \texttt{gwsubtract} data is systematically biased, which is the reason why \citet{Payne_2022} also used an alternative method to clean the Livingston data. 
Ref.\,\cite{Payne_2022} mitigates the excess noise with BayesWave, an algorithm that does not use witness channels and relies only on the strain data to model the noise~\citep{bw1, bw2}. 
Their analysis found that the evidence for precession is smaller than the statistical and systematic uncertainty of noise mitigation.

In this paper, we describe a novel method to mitigate the excess noise using machine learning, with a focus on removing the noise around GW200129. 
In \S\ref{sec:methods}, we describe our data selection procedure, the machine learning model, how the effectiveness of the model is tested, as well as the procedure to estimate the orbital precession for GW200129.
In \S\ref{sec:results}, we show how well the machine learning model works, compare its effectiveness against \texttt{gwsubtract} and present results of the spin-induced orbital precession estimate.
We summarize our findings in \S\ref{sec:conclusions}.

\section{Methods}
\label{sec:methods}
\subsection{Data}%
\begin{figure}[htpb]
    \centering
    \includegraphics[width=\columnwidth]{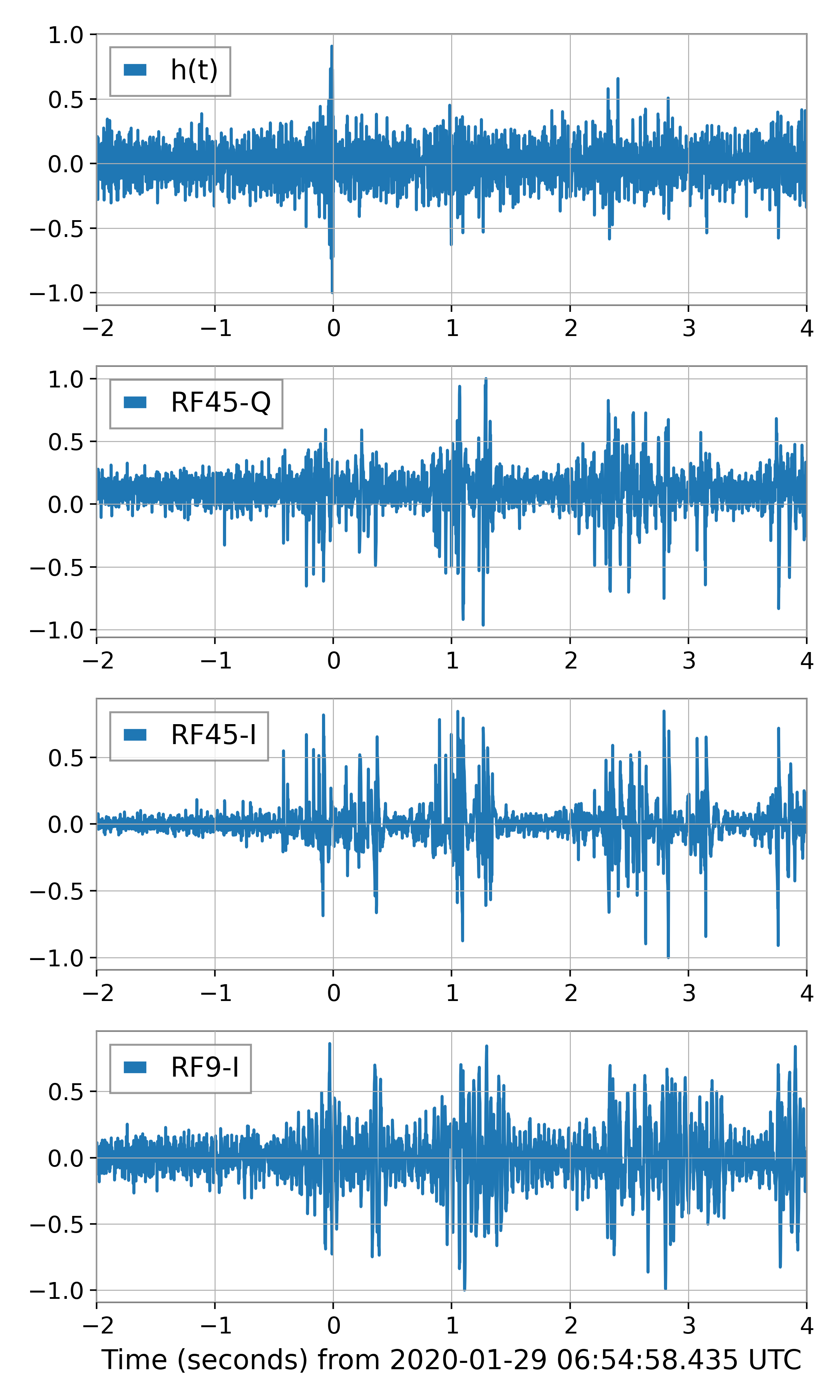}
    \caption{
             Whitened time series data for the gravitational-wave strain channel \texttt{h(t)} and witness channels \texttt{RF45-Q}, \texttt{RF45-I} and \texttt{RF9-I}.
             Gravitational-wave signal GW200129 is clearly seen in \texttt{h(t)} at $0$\,s.
             The non-Gaussian noise in \texttt{h(t)} caused by the electro-optic modulator system is also recorded by the witness channels.
             }
    \label{fig:tseries}
\end{figure}

\label{sub:methods_data}
The Livingston interferometer was locked and observing for more than five hours when GW200129 was detected and continued to observe for more than 38 hours.
During this time, the Livingston interferometer had excess noise caused by the $45$\,MHz electro-optic modulator system; this noise is also known as the radio frequency (RF) noise~\cite{rf_noise}.
RF noise coupled to the GW strain channel \texttt{h(t)}, as well as the noise witness channels \texttt{RF45-I}, \texttt{RF45-Q} and \texttt{RF9-I} (Figure \ref{fig:tseries}).

To learn the coupling between the witness channels and the GW strain channel \texttt{h(t)}, we selected 27 hours of the total 43 hours when the detector was locked and observing.
To be exact, we chose the interval from Jan 29, 2020 09:55:06 UTC to Jan 30, 2020 12:55:06 UTC. 

We chose this interval due to multiple reasons.
Firstly, the RF noise was exceptionally recurrent during this time.
In addition, the noise coupling may change if the interferometer configuration changes, therefore we selected only the data contained in this observational lock.
Finally, we did not use the time around GW200129 to avoid any biases.

After data selection, we whitened the data using the inner product equation, where $\braket{a|b}$ is the inner product between the two time-series $a(t)$ and $b(t)$ given by
\begin{equation}
    \label{eq:inner_product}
    \braket{a|b} = 4 \int_0^\infty \frac{\tilde{a}(f)\tilde{b}^*(f)}{S_n(f)}df,
\end{equation}
where $S_n(f)$ denotes the power spectral density (PSD).
To avoid using PSD biased by the RF noise, we inspected the data visually and selected $512$\,s of relatively quiet data for the PSD estimation.

Once the data was whitened, it was resampled to $512$\,Hz.
Because our network memory requirement is proportional to $f^{3}_\textrm{Nyquist}$, data resampling from $4096$\,Hz to $512$\,Hz significantly reduced the required memory to train the algorithm.
Since the orbital precession for GW200129 is mostly observed below $50$\,Hz~\cite{Payne_2022}, reducing the sampling rate does not impact the precession measurement.

After resampling the data, we selected the time intervals when the RF noise was present in the witness channels.
To identify these intervals, the Z-score of $11.7$ was used for the \texttt{RF45-I} channel (there was no particular reason why this specific channel was selected out of the three witness channels).
If a data point passed this threshold, we would select $4$\,s around this data point.

We collated $477$ noisy data intervals (about 32 minutes) after this procedure.
Using $\sim$32 minutes of noise-only data instead of the full 27 hours of data allowed us to train the machine learning algorithm more efficiently.

\subsection{Dense Neural Network}%
\label{sub:methods_ml}

We use a Dense Neural Network (DNN) to model the excess noise witnessed by the RF channels.
DNNs are well known for their ability to generalize even extremely complicated relationships in data, including the non-linear coupling which can happen among multiple noise witness channels.
In addition, a DNN does not rely on the assumption that the data needs to be Gaussian and stationary like the linear subtraction method \texttt{gwsubtract}.

We found that a relatively simple fully connected DNN with just two hidden layers and a dropout layer in between is enough to model the RF noise around GW200129 (see Table\,\ref{tab:arch} for full network architecture).
To predict a single noise point, i.e.\,$\frac{1}{512}$\,s, in \texttt{h(t)}, we use $2$\,s of data from each RF channel.

In order to reduce the overfitting and increase the learning efficiency, we use L2 kernel regularization of $10^{-4}$ and ReLu activation function.
The dataset from \S\ref{sub:methods_data} is split with 9:1 ratio for testing and validating the algorithm.
The network is trained for $500$ epochs using Adam optimizer~\cite{adam} with inverse time decay learning function, learning rate of $10^{-3}$ and Mean Squared Error loss function.
Training the network written with \texttt{TensorFlow}~\cite{tf} takes about $30$ minutes on Nvidia A100 80GB GPU.
\begin{table}
\centering
\small
    \begin{tabular}{c c}
        \hline
        Layer & Output Shape (length, dimensions)\\ [0.5ex]
        \hline\hline
        Input & (1024, 3) \\
        Flatten & (3072, 1) \\
        Dense & (1024, 1) \\
        Dropout & (1024, 1) \\
        Dense & (1024, 1) \\
        Output & (1, 1) \\
        \hline
        \hline
    \end{tabular}\caption{\label{tab:arch} Dense neural network architecture. We use $2$\,s of data with $f_\textrm{Nyquist}=256$\,Hz from each RF channel to predict a single data point, i.e.\,$\frac{1}{512}$\,s, in the gravitational-wave strain channel \texttt{h(t)}.}
\end{table}

After the training is performed, the model can be applied to mitigate the noise around GW200129.
To do that, we perform the same steps as in \S\ref{sub:methods_data} for $4096$\,s around GW200129, i.e.\,the data is firstly whitened and resampled. 
After that, the three RF channels are used as input to get the RF noise contribution in \texttt{h(t)} estimated by the DNN.
Once that is done, the data is upsampled to $4096$\,Hz, de-whitened and subtracted from the original $4096$\,s LIGO Livingston frame.
This is the data frame which has RF noise mitigated using our machine learning algorithm.
The data frame is publicly available on Zenodo\,\footnote{DOI: 10.5281/zenodo.10143337}.

\subsection{Testing the cleaned data}
\label{sub:methods_pymc}
We apply the method proposed by Macas \& Lundgren (2023) to estimate how much of the excess noise is removed using our algorithm, and compare its effectiveness with \texttt{gwsubtract}~\cite{pymc_stats}.
To do that, we firstly identify time intervals when the RF noise is present in the data around GW200129.
Using the Z-score of 6 for the \texttt{RF45-I} channel, we select $366$\,s out of $4096$\,s considered to contain the radio frequency noise.

We convert this data into the normalised time-frequency tiles with quality factor $Q=8$ which allows us to estimate the average tile power.
Then, using Bayesian statistical modelling, we calculate the amount of power in the non-Gaussian data versus the total power, defined as \textit{fractional power}.
We repeat the same procedure for the original Livingston data, as well as the data that has the RF noise mitigated with \texttt{gwsubtract}.

\subsection{Estimating the spin-induced orbital precession of GW200129}
\label{sub:methods_pe}
We finally apply the \texttt{bilby} Bayesian inference package~\citep{ashton_2021,romero_2020} to estimate the posterior probability density using our new cleaned Livingston data and the original data from Hanford and Virgo.
We follow the configuration of the initial LVK GWTC-3 analysis \citep{o3b_cat}, but use the \texttt{NRSur7dq4} waveform approximant~\citep{Varma:2019csw} with a suitably modified prior following \citet{Hannam_2022}.
We also marginalise over the calibration uncertainty.
However, the PSD used by the GWTC-3 analysis, used an on-source estimate which depends on the analysis data itself.
Since we have cleaned the Livingston data, we opt instead to use a Welch PSD estimate for all three detectors.
Finally, we perform sampling using the \texttt{bilby-MCMC} sampler~\citep{ashton_2021}.

\section{Results and Discussion}
\label{sec:results}

\subsection{Data cleaning}%
\begin{figure}[htpb]
    \centering
    \includegraphics[width=0.4\paperwidth, clip=true, trim={0 0 0 2cm}]{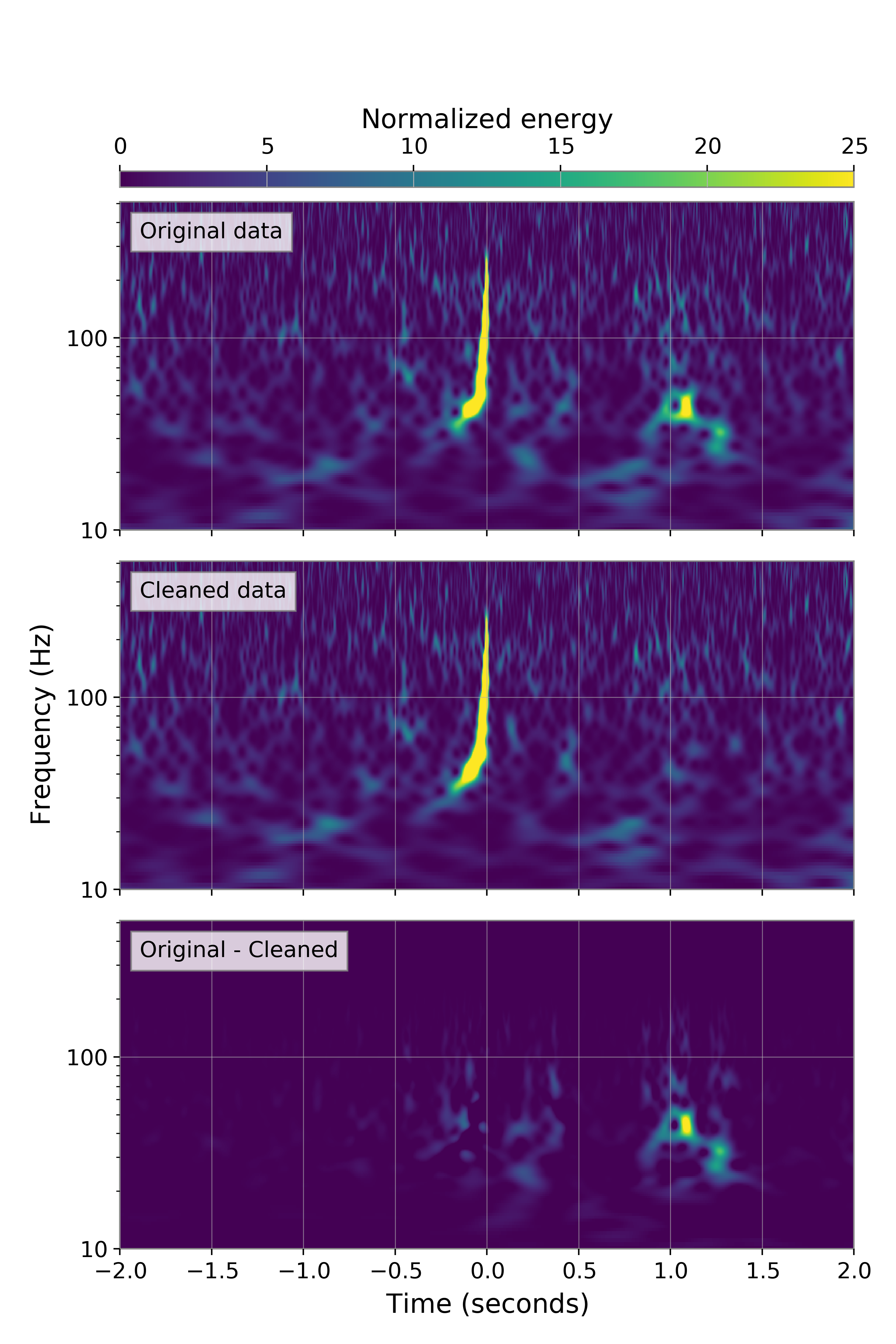}
    \caption{
             Spectrograms of the original Livingston data around GW200129, cleaned Livingston data using our machine learning approach, and the difference between these two spectrograms.
             Our algorithm removes the non-Gaussian noise after the GW signal, as well as some noise that happened during the GW signal and within the frequency range of the GW signal.
             }
    \label{fig:oscan}
\end{figure}

\label{sub:data_cleaning}
Figure~\ref{fig:oscan} shows the spectrograms of the Livingston data around GW200129 before cleaning, after cleaning and the difference between these two spectrograms.
We note that the algorithm removed large portions of the excess noise after the GW signal, as well as some noise that happened during the GW signal and within the frequency range of the GW signal.
Comparing with the \texttt{gwsubtract} results~\cite{Davis_2022}, it is clear that our algorithm removes noticeably more data at frequencies above $50$\,Hz.

\begin{figure}[htpb]
    \centering
    \includegraphics[width=\columnwidth]{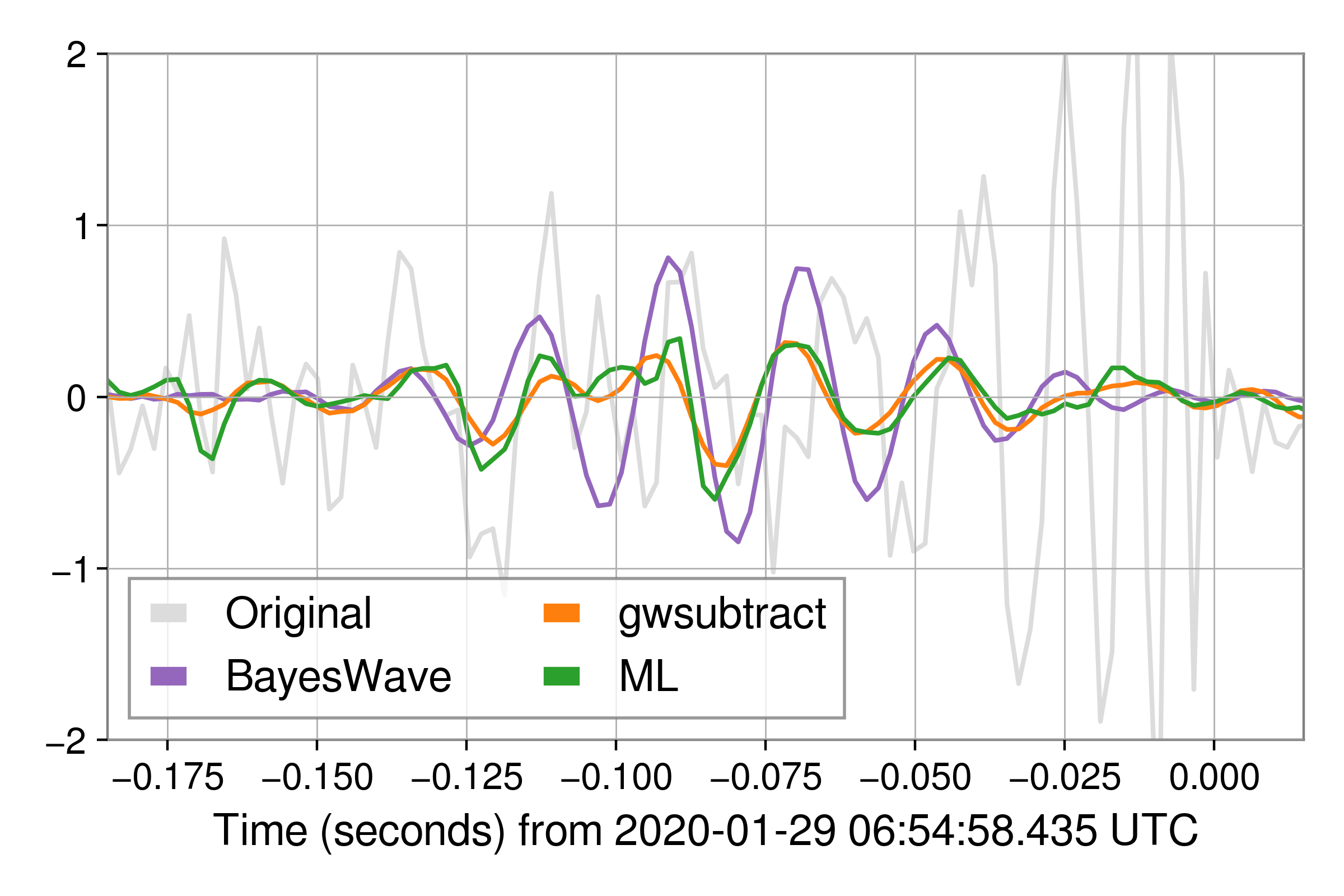}
    \caption{
             Radio frequency noise modelled by BayesWave (purple), gwsubtract (orange) and our algorithm (green); the original unmitigated data is in gray.
             All data was whitened and bandpassed to $512$\,Hz.
             }
    \label{fig:noise_models}
\end{figure}

As we can see from Fig.~\ref{fig:noise_models}, our modelled radio frequency noise is closer to the \texttt{gwsubtract} noise prediction than the prediction made by BayesWave (glitch model A) in \citet{Payne_2022}.
This makes sense since both, \texttt{gwsubtract} and ML approach use witness channel information.
However, ML subtraction contains more high-frequency features than the \texttt{gwsubtract}, which can be explained by the fact that ML uses more witness channels and/or the algorithm represents the non-linearity better.
Comparing BayesWave's noise model with \texttt{gwsubtract} and ML, it looks much simpler with a peak around $-0.08$\,s before the merger.

Close to the merger time, we can see some differences between all three models.
Around $-0.025$\,s before the merger, BayesWave subtracts more data than \texttt{gwsubtract} and ML.
Later, around $-0.015$\,s before the merger, \texttt{gwsubtract} subtracts more data than BayesWave but less data than ML.

\begin{figure}
    \centering
    \begin{subfigure}[b]{0.4\paperwidth}
        \centering
        \includegraphics[width=0.4\paperwidth]{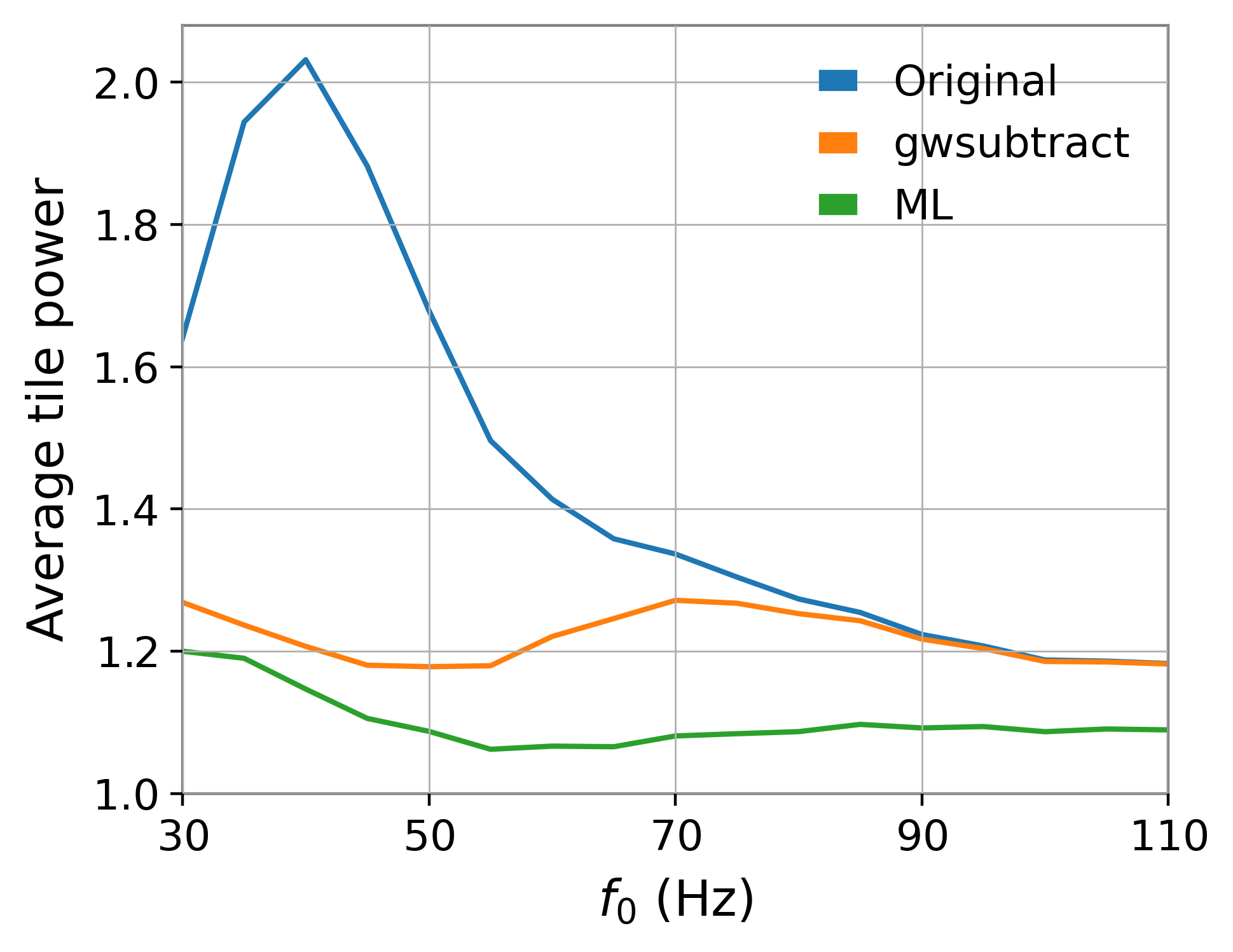}
        \caption{Average tile power for data around GW200129 with quality factor $Q=8$.}\label{fig:tpower}
    \end{subfigure}
    \begin{subfigure}[b]{0.4\paperwidth}
        \centering
        \includegraphics[width=0.4\paperwidth]{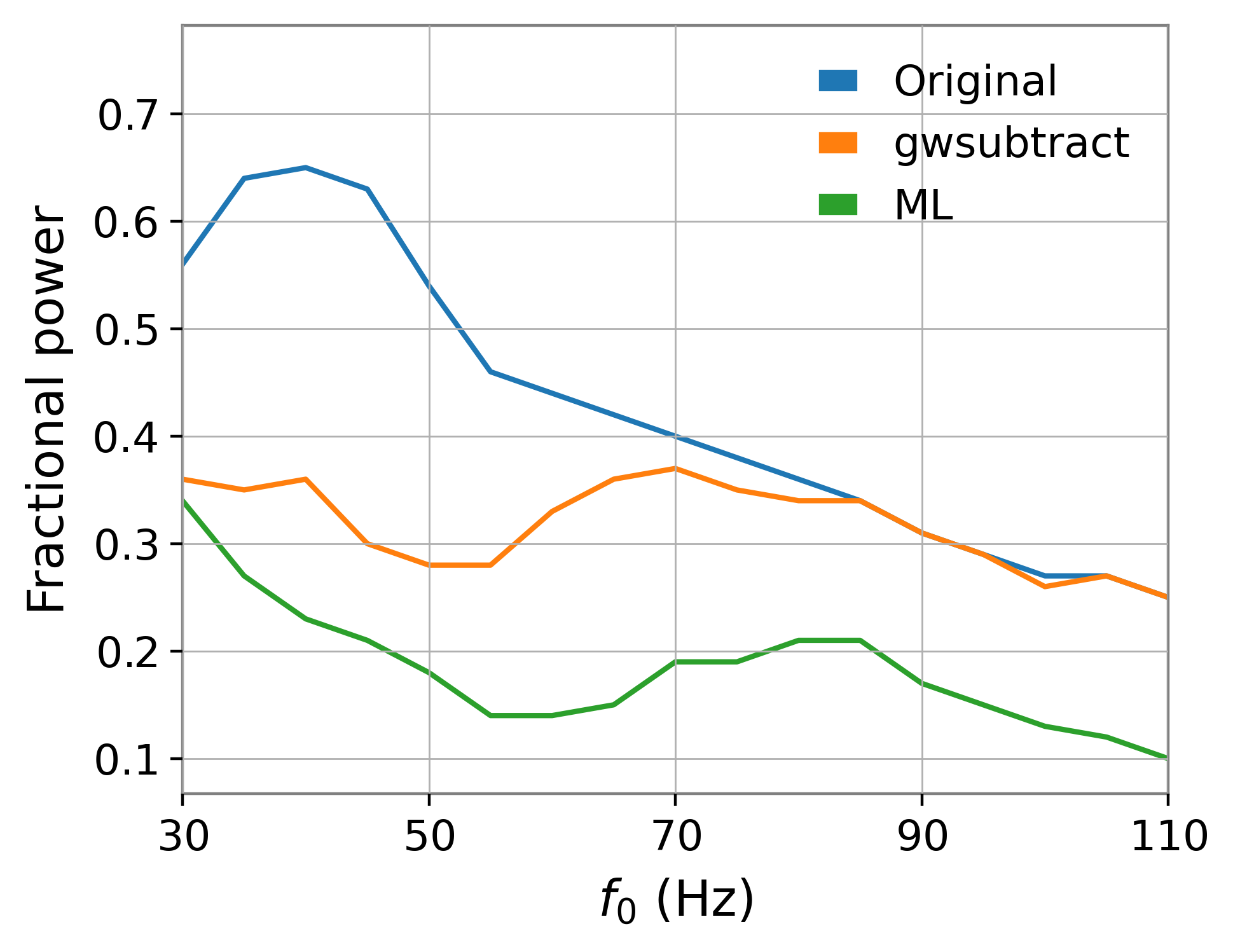}
        \caption{Fractional power for data around GW200129 with quality factor $Q=8$.}\label{fig:fpower}
    \end{subfigure}
    \caption{
             Average tile power (above) and fractional power (below) for data around GW200129 with quality factor $Q=8$.
             Both tests indicate that our machine learning algorithm removes more non-Gaussian noise than \texttt{gwsubtract}, especially at frequencies above $50$\,Hz. 
             }
    \label{fig:powers}
\end{figure}

For a more quantitative comparison between the original, \texttt{gwsubtract} and ML data frames, see Figure~\ref{fig:powers}.
We omit results of \citet{Payne_2022} in this comparison because the \citet{pymc_stats} comparison method requires significantly more data than the Ref.~\cite{Payne_2022} cleaned.

The average tile power plot (Fig.~\ref{fig:tpower}) shows that both \texttt{gwsubtract} and ML algorithms significantly reduce the excess power.
For Gaussian data, the average tile power is 1, while the original data frame has as much as twice the amount of power than Gaussian data around $40$\,Hz.

ML algorithm removes more excess power than \texttt{gwsubtract} across all frequency ranges, and the difference becomes much higher at frequencies above $\sim 50$\,Hz.
From $\sim85$\,Hz, \texttt{gwsubtract} does not remove any excess noise any more.
This happens because \texttt{gwsubtract} uses only a single witness channel that does not contain any correlation with \texttt{h(t)} at those frequencies.

The fractional power plot (Fig.~\ref{fig:fpower}) shows similar conclusions.
ML algorithm is better at removing the excess noise across all frequencies, reducing up to 10\% more of the fractional power at frequencies below $50$\,Hz.
At higher frequencies, the difference in fractional power between \texttt{gwsubtract} and ML can be as high as 20\%.

Unfortunately, the average tile power and fractional power plots (Figs.~\ref{fig:tpower} and \ref{fig:fpower}) indicate that neither the \texttt{gwsubtract} nor the ML algorithm remove the excess noise completely.
This can happen due to a variety of reasons.
Firstly, it is possible that both algorithms cannot estimate the correlation function between the witness channels and the gravitational-wave strain ideally.
Another possibility is that none of the witness channels are perfect, i.e.\,they do not record the full coherence between the RF noise in a witness channel and the RF noise in \texttt{h(t)}.

\subsection{Spin-induced orbital precession of GW200129}%
\label{sub:pe_results}

\begin{figure}
\centering
\includegraphics[width=0.4\paperwidth,trim={0 2cm 0 2cm},clip]{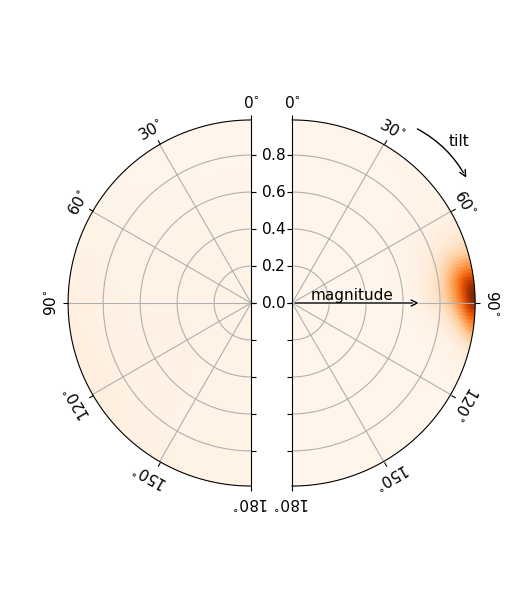}
\caption{A spin-disk plot showing the two-dimensional posterior distribution of the spin magnitude and tilt of the primary (right-hand panel) and secondary (left-hand panel) of GW200129.}
\label{fig:spin}
\end{figure}

In Fig.~\ref{fig:spin}, we show the posterior distribution of the spin magnitude and tilt from our new analysis utilising the cleaned data.
Comparing with Fig.~2 of \citet{Hannam_2022} which used the \texttt{gwsubtract} cleaned frame, we see similar results within the expected sampling error (any subtle changes would most likely be caused by the difference in PSD construction).
From this figure, we can confirm that the identification of evidence for precession, i.e.\,that the primary spin is constrained to be close to unity and highly misaligned, and is robust to our new de-glitching routine.

\begin{figure}
    \centering
    \includegraphics[width=0.4\paperwidth]{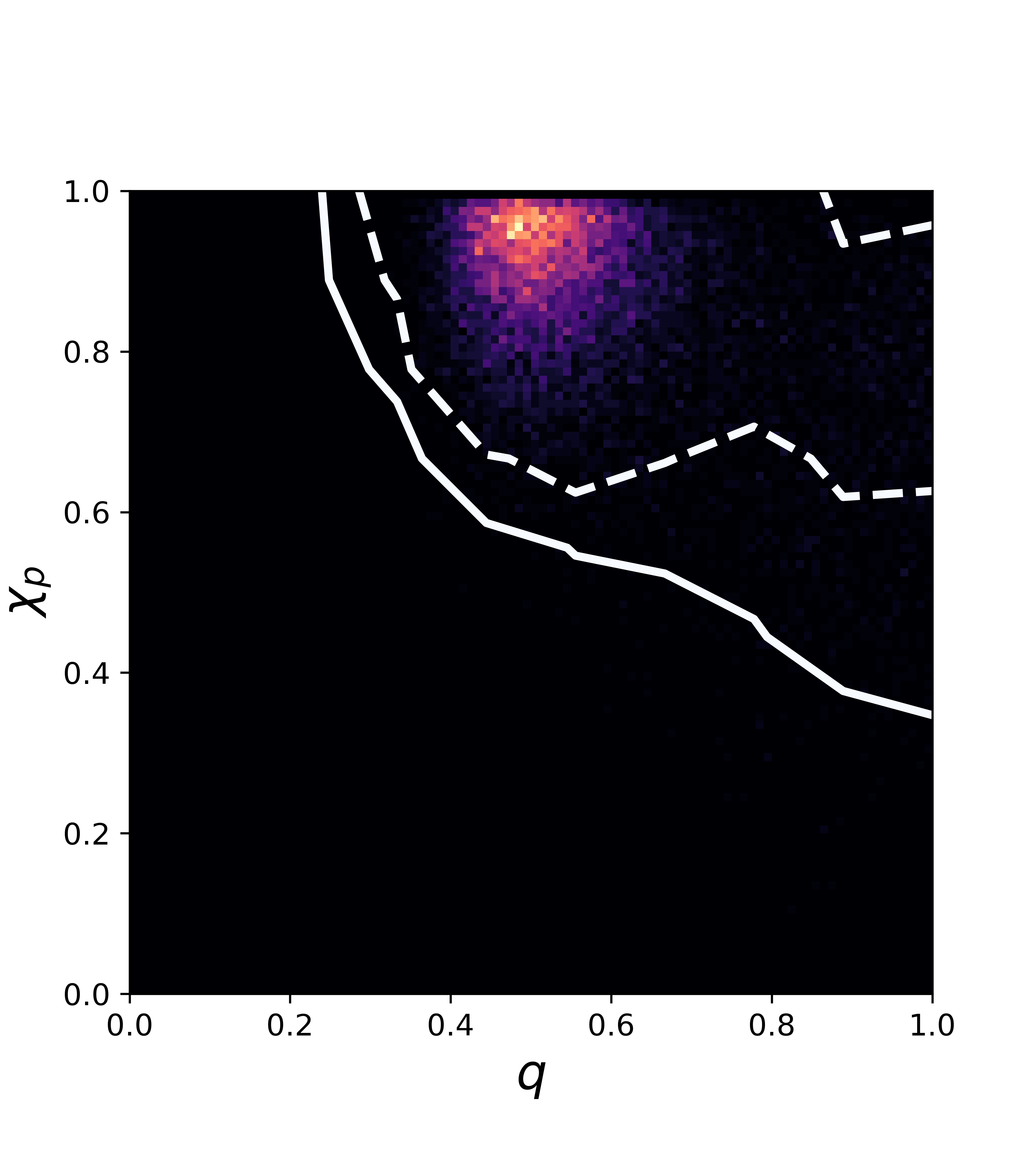}
    \caption{Mass ratio $q$ and the precession parameter $\chi_p$ plot for GW200129 using ML-mitigated data.}
    \label{fig:chi_p}
\end{figure}

In Fig.\ref{fig:chi_p}, we demonstrate the two-dimensional mass ratio $q$ and precession parameter $\chi_p$ results using our noise mitigation routine.
We see that the mass ratio peaks around $0.5$ and that $\chi_p$ peaks around $0.95$ indicating a highly precessing system.
This agrees with results from \citet{Hannam_2022} but contrasts with the findings from \citet{Payne_2022} where BayesWave is used to model the noise.
\citet{Payne_2022} finds the mass ratio to be less symmetric and the posterior on the precession parameter $\chi_p$ to be less informative (depending on the glitch model they use).

\section{Conclusions}
\label{sec:conclusions}
In this paper, we demonstrated that machine learning can model the non-Gaussian noise recorded by witness channels. 
We use \texttt{RF45-I}, \texttt{RF45-Q} and \texttt{RF9-I} channels as input to the fully connected dense neural network and gravitational-wave strain channel \texttt{h(t)} as output. 
Such an algorithm allowed us to mitigate the noise around the GW200129 signal.

We also found that our approach removes a significant part of the non-Gaussian noise in \texttt{h(t)}, and that it actually removes more non-Gaussian noise than \texttt{gwsubtract}, the algorithm used by the LIGO-Virgo-KAGRA collaboration.

At frequencies below $50$\,Hz, our algorithm removes up to 10\% more fractional power compared with \texttt{gwsubtract}, while at higher frequencies the fractional power difference can be as high as 20\%.
However, neither \texttt{gwsubtract} nor our method is able to remove the non-Gaussian noise completely.
This can happen due to these algorithms being imperfect, witness channels not containing the full information about the excess noise, or any other unknown reason.

Furthermore, we note that our method does not over-subtract the noise.
The algorithm removes only the correlated noise between the gravitational-wave strain $h(t)$ and the witness channels as the algorithm does not have enough capacity to memorize all of the data it is trained on.

The re-analysis of GW200129 with the improved data quality finds a similar evidence of the spin-induced precession as reported by \citet{Hannam_2022}.
However, we note that while our method is able to remove more RF noise from the data than \texttt{gwsubtract}, we also show that there is still some noise left unmitigated.

\acknowledgments
We thank Adrian Helmling-Cornell, Charlie Hoy, Christopher Berry, Katerina Chatziioannou, Mark Hannam, Nikolaos Stergioulas, Thomas Dent and members of the LIGO Detector Characterization group for their valuable input during the preparation of this manuscript.
We also thank the anonymous referee for valuable comments which improved the manuscript during the review.
RM is supported by STFC grants ST/S000550/1, ST/T000325/1, ST/V005715/1 and ST/X002225/1.
The authors are grateful for computational resources provided by the LIGO Laboratory and supported by National Science Foundation Grants PHY-0757058 and PHY-0823459.
This research has made use of data, software and/or web tools obtained from the Gravitational Wave Open Science Center (https://www.gw-openscience.org), a service of LIGO Laboratory, the LIGO Scientific Collaboration, the Virgo Collaboration, and KAGRA. This material is based upon work supported by NSF’s LIGO Laboratory which is a major facility fully funded by the National Science Foundation. LIGO Laboratory and Advanced LIGO are funded by the United States National Science Foundation (NSF) as well as the Science and Technology Facilities Council (STFC) of the United Kingdom, the Max-Planck-Society (MPS), and the State of Niedersachsen/Germany for support of the construction of Advanced LIGO and construction and operation of the GEO600 detector. Additional support for Advanced LIGO was provided by the Australian Research Council. Virgo is funded, through the European Gravitational Observatory (EGO), by the French Centre National de Recherche Scientifique (CNRS), the Italian Istituto Nazionale di Fisica Nucleare (INFN) and the Dutch Nikhef, with contributions by institutions from Belgium, Germany, Greece, Hungary, Ireland, Japan, Monaco, Poland, Portugal, Spain. KAGRA is supported by Ministry of Education, Culture, Sports, Science and Technology (MEXT), Japan Society for the Promotion of Science (JSPS) in Japan; National Research Foundation (NRF) and Ministry of Science and ICT (MSIT) in Korea; Academia Sinica (AS) and National Science and Technology Council (NSTC) in Taiwan.
Various parts of the analysis used \texttt{GWpy} \cite{gwpy}, \texttt{PyCBC} \citep{pycbc, pycbc_live_o3}, \texttt{bilby} \citep{ashton_2021, romero_2020}, \texttt{bilby-mcmc} \cite{ashton_2019}, \texttt{TensorFlow} \cite{tf}, \texttt{Keras} \cite{keras}, \texttt{Adam} \cite{adam}, \texttt{PyMC} \cite{pymc}, \texttt{NumPy} \cite{numpy}, \texttt{SciPy} \cite{scipy}, \texttt{IPython} \cite{ipython}, \texttt{Jupyter notebook} \cite{jupyter}, \texttt{pandas} \cite{pandas}, \texttt{Matplotlib} \cite{matplotlib}, \texttt{arviz} \cite{arviz} and \texttt{corner} \cite{corner}.
This document has been assigned LIGO Laboratory document number LIGO-P2300358.

\clearpage

\bibliography{references, software}

\end{document}